\documentclass[12pt]{article} 
\usepackage{jheppub}

\usepackage{amsmath, amsfonts, amssymb}
\usepackage{mathrsfs}

\usepackage{graphicx,epsfig}
\usepackage{epic}
\usepackage{color}

\newcommand{\be}{\begin{equation}}
\newcommand{\ee}{\end{equation}}
\newcommand{\ba}{\begin{eqnarray}}
\newcommand{\ea}{\end{eqnarray}}

\newcommand{\ads}{$AdS_5\times S^5$\ }

\newcommand{\adscp}{$AdS_{4}\times \mathbb{CP}^{3}$}
\newcommand{\maldafive}{${\rm AdS}_{5}/{\rm CFT}_{4}$\ }

\newcommand{\mc}{\mathcal }

\def\XXint#1#2#3{{\setbox0=\hbox{$#1{#2#3}{\int}$}
     \vcenter{\hbox{$#2#3$}}\kern-.5\wd0}}

    \newcommand{\beq}{\begin{equation}}
    \newcommand{\eeq}{\end{equation}}
    \newcommand\beqa{\begin{eqnarray}}
    \newcommand\eeqa{\end{eqnarray}}

\title{Semiclassical folded string in \adscp}

\author[a]{Matteo Beccaria } 
\author[b]{, Guido Macorini} 
\author[a]{, CarloAlberto Ratti} 
\author[c]{, Saulius Valatka} 

\affiliation[a]{Dipartimento di Matematica e Fisica ``Ennio De Giorgi``, Universita' del Salento \& INFN, \\
                     Via Arnesano, 73100 Lecce, Italy} 
                     
\affiliation[b]{Niels Bohr International Academy and Discovery Center,  
		     Niels Bohr institute, \\
		     Blegdamsvej 17 DK-2100 Copenhagen, Denmark}
		
\affiliation[c]{Mathematics Department, King’s College London, \\
	The Strand, London WC2R 2LS, UK}

\emailAdd{matteo.beccaria$\bullet$le.infn.it}
\emailAdd{macorini$\bullet$nbi.ku.dk}
\emailAdd{carloalberto.ratti$\bullet$le.infn.it}
\emailAdd{saulius.valatka$\bullet$kcl.ac.uk}

\abstract{
We consider type IIA superstring theory on the background \adscp\, and the classical solution describing a folded string 
spinning in $AdS_{4}$  with angular momentum in $\mathbb{CP}^{3}$. In the 't Hooft limit, it is the gravity dual
of twist operators in the ABJM superconformal theory. We quantize the classical solution by algebraic curve methods and 
determine the first semiclassical correction to the energy. We provide 
an integral representation for this quantity valid for all values of the charges. We analyze its properties in the special regimes
associated with a short or long string providing various accurate analytical expansions. Finally, we investigate 
various properties of the so-called slope, the leading term of the energy for short strings, collecting  information 
that could be useful in attempts to generalize the exact results recently proposed for the folded string in \ads.
 }

\arxivnumber{1234.5678}

\begin{document} \maketitle

\bigskip

\section{Introduction}

The AdS/CFT correspondence \cite{Maldacena:1997re,Witten:1998qj,Gubser:1998bc} is an extremely deep and fruitful theoretical idea that played a central role in the past decade. 
In its simplest instance, it links the integrability of the world-sheet $\sigma$-model for type IIB superstring
to the strong coupling behaviour of four dimensional gauge theories. In the specific case of planar $\mc N=4$
Super Yang-Mills theory, integrability methods determine the spectrum of anomalous dimensions as a function of 
the 't Hooft coupling $\lambda$, from weak to strong coupling.

After the foundational papers applying integrability methods in AdS/CFT 
\cite{Minahan:2002ve,Bena:2003wd,Arutyunov:2004vx,Beisert:2005tm,Janik:2006dc,Beisert:2006ib,Beisert:2006ez}
(see also the recent review \cite{Beisert:2010jr}), a solution to the spectral problem was  obtained for asymptotically long single trace operators by means of the Bethe Ansatz approach \cite{Beisert:2005fw}. 
Later, the full spectrum was shown to be captured by suitable Y-system equations 
\cite{Gromov:2009tv,Arutyunov:2009zu,Bombardelli:2009ns,Gromov:2009bc,Arutyunov:2009ur}. These equations
are expected to be general and to be valid for any operator, including short ones.

The Y-system equations are definitely involved, but in \cite{Gromov:2009bc} they had been formulated 
for the single trace $\mathfrak{sl}(2)$ twist operators in a form suitable for  numerical studies.
This opened the way to explorations of the flow of the anomalous dimension of the most famous among them, the Konishi operator. This investigation culminated in the analysis of \cite{ Gromov:2009zb} where the $\lambda$ dependance has been studied  interpolating from the weak coupling expansion, known explicitly from perturbative calculations in gauge theory up to four loops, to the strong coupling string theory prediction \cite{Gubser:2002tv} known only to tree level.

The next logical step was the determination of the analytical strong coupling expansion of the Konishi anomalous dimension, {\em i.e.} the energy of its gravity dual state. The sub-leading coefficient in the string theory expansion was found independently by three groups 
\cite{Gromov:2011de,Roiban:2011fe,Vallilo:2011fj} (see also \cite{Beccaria:2011uz} for related work on a different superconformal representative of Konishi). In the end, the numerical predictions of \cite{Gromov:2009zb} have been confirmed. More recently a highly nontrivial observation was made in \cite{Basso:2011rs}, which allows one to reproduce the one-loop result almost without any effort. Finally, the second nontrivial strong coupling expansion coefficient was derived analytically in \cite{Gromov:2011bz} by using the one-loop expression for a 
general  $(S,J)$ folded string \cite{Frolov:2002av} found in  \cite{Gromov:2011de} and the conjecture proposed 
in \cite{Basso:2011rs}. The comparison with the available numerical data from the thermodynamical Bethe Ansatz approach shows a rather promising agreement. 

\bigskip
The folded string is indeed a very useful theoretical laboratory. At the classical level, it is convenient to describe it in terms of the scaled charges $\mc S = S/\sqrt\lambda$, $\mc J = J/\sqrt\lambda$. They have a simple and clear
dynamical meaning. The spin $\mc S$ is associated with rotation around the center of 
$AdS_{5}$ while $\mc J$ is an angular momentum on the sphere. At small $\mc S$, the string is short and admits
a near-flat space description. At large $\mc S$, the string stretches and reaches the boundary of $AdS$ with 
the characteristic scaling $E \sim \log \mc S$ of its energy. Quantum corrections in the semiclassical approximation
are fully under control and can be studied by algebraic curve tools. Also, the folded string is a clear example where integrability techniques are quite competitive with traditional field-theoretical 
methods \cite{Beccaria:2010ry,Beccaria:2008tg} (see also \cite{Beccaria:2012mx} for a similar approach to 
the more complicated case of spiky strings).

\bigskip
Remarkably, very similar integrability structures as well as folded string classical solutions are known to be 
available in the ABJM theory~\cite{Aharony:2008ug}. This is a three dimensional $U(N)\times U(N)$ gauge theory with four complex scalars in the $(N, \overline N)$ representation, 
their fermionic partners, and a Chern-Simons action with levels $+k$, $-k$. This theory has ${\cal N}=6$ superconformal symmetry $\mathfrak{osp}(2,2|6)$.
ABJM can be considered as the low energy theory of $N$ parallel M2-branes at a $\mathbb{C}^4/\mathbb{Z}_k$ singularity. In the large $N$ 
limit this is M theory on $AdS_4\times S^7/\mathbb{Z}_k$. For fixed $\lambda=N/k$ we can describe it by type IIA string on $AdS_4\times \mathbb{CP}^3$
which is classically integrable~\cite{Arutyunov:2008if,Stefanski:2008ik,Gomis:2008jt}.
The manifest (non abelian) part of the R symmetry is $SU(2)\times SU(2)$. The complex scalars can be written as two  doublets transforming as $(2,1)$ and $(1,2)$.
Under the gauge group they transform as $(N, \overline N)$ and $(\overline N, N)$.
At leading order (two loops, $\lambda^2$ in 't Hooft coupling $\lambda$), the dilatation operator for single trace operators built with 
these scalars is a $SU(4)$ integrable spin chain~\cite{Minahan:2008hf,Bak:2008cp}.
In~\cite{Gromov:2008qe}, Gromov and Vieira have proposed a set of all-loop Bethe-Ansatz equations for the full $\mathfrak{osp}(2,2|6)$ theory consistent with the two loops analysis and with the 
superstring algebraic curve at strong coupling~\cite{Gromov:2008bz}. The equations depend on a dressed coupling $h(\lambda)$ which takes into account the fact that 
the one-magnon dispersion relation is not protected~\cite{Nishioka:2008gz,Gaiotto:2008cg,Grignani:2008is,McLoughlin:2008he}.
Finally,  an exact  $SU(2|2)_A \oplus SU(2|2)_B$ symmetric $S$-matrix consistent with~\cite{Gromov:2008qe} has been presented in~\cite{Ahn:2008aa}. A comprehensive review of integrability methods in ABJM discussing also their available checks can be found 
in \cite{Klose:2010ki}.

The important point for our investigation is that there is a  $\mathfrak{sl}(2)$-like sector in ABJM as discussed in~\cite{Gromov:2008qe,Zwiebel:2009vb}.
At strong coupling and large spin these operators behave quite similarly to the corresponding ones in \ads. In particular their dual string state is a folded string 
rotating in $AdS_3$  and with angular momentum $J$ in 
$\mathbb{CP}^3$~\cite{McLoughlin:2008ms,Alday:2008ut,Krishnan:2008zs}. 
At weak coupling, they are composite operators in totally different theories. Nevertheless, 
both ${\cal N}=6$ Super Chern-Simons and ${\cal N}=4$ Super Yang-Mills are integrable and the all-loop Bethe equations in the $\mathfrak{sl}(2)$ sectors are essentially the same~\footnote{See 
\cite{Beccaria:2009ny} for higher loop calculations in specific short states.}. 

\bigskip
The current treatment of quantized folded string in \adscp\ \cite{Chen:2008qq} is still rather limited. The large spin limit has 
been studied at leading order in various regimes: (a) $J/S$ fixed  in \cite{McLoughlin:2008he}, (b) 
 $J\sim \log S$ in \cite{Gromov:2008fy}. The short string limit has never been addressed. On the numerical side, 
 the analysis of \cite{LevkovichMaslyuk:2011ty} explored the TBA equations of \adscp, although in a small window of the 
 coupling reaching at most $\lambda \sim \mc O(1)$.
 In this paper, we present a complete analysis of the one-loop semiclassical energy of folded string in \adscp.  In more details, 
 \begin{enumerate}
 \item[(a)]
 we derive a simple and compact integral representation for the one-loop energy correction valid for generic values of the charges.
 
 \item[(b)]  At small (semiclassical) spin, we discuss the explicit evaluation of the sum over frequencies
 and obtain various short string expansions like that in \cite{Gromov:2011bz} or the resummed one discussed in \cite{Beccaria:2012tu}.  
 
 \item[(c)] At large spin, we extract from the integral representation the energy expansion at next-to-leading order.
 
 \item[(d)] Finally, we discuss 
 in some detail the structure of the
 {\em slope} in ABJM, {\em i.e.} the linear term in the small spin expansion of the one-loop energy as a first step 
toward a generalized Basso's conjecture \cite{Basso:2011rs} for this theory. 

\end{enumerate}

\section{Algebraic curve quantization in \adscp}

We give in this section a compact self-contained summary of the results of \cite{Gromov:2008bz} using the 
language of off-shell fluctuation energies \cite{Gromov:2008ec}. We shall work in the algebraic curve 
regularization and write all equations in terms of the $\sigma$-model coupling $g$. For large $g$, it is 
related to the 't Hooft coupling by 
\be
\lambda = N/k = 8\,g^{2},
\ee
but, contrary to the \ads\ case, this relation will get corrections at finite $g$.
The classical algebraic curve for \adscp\ is a 10-sheeted Riemann surface. The spectral parameter moves on it and 
we shall consider 10 symmetric quasi momenta $q_{i}(x)$ 
\be
(q_{1}, q_{2}, q_{3}, q_{4}, q_{5}) = (-q_{10}, -q_{9}, -q_{8}, -q_{7}, -q_{6}).
\ee
They can have branch cuts connecting the sheets with 
\be
q_{i}^{+}-q_{j}^{-} = 2\,\pi\,n_{ij}.
\ee
In the terminology of \cite{Gromov:2008bz}, the physical polarizations $(ij)$ can be split into {\em heavy} and {\em light} ones and are summarized in the following table:
 $$
 \begin{array}{c|ccc}
  & \mbox{AdS${}_{4}$} & \mbox{Fermions} & \mathbb{CP}^{3} \\
  \hline
  \mbox{heavy} & \quad (1,10) (2,9) (1,9)\quad & (1,7) (1,8) (2,7) (2,8) & (3,7) \\
  \mbox{light} & & (1,5) (1,6) (2,5) (2,6) & \quad (3,5) (3,6) (4,5) (4,6)\quad
  \end{array}
 $$
 Virasoro constraints require that the poles of the quasi-momenta $q_{i}(x)$ at $x=\pm 1$ are synchronized according to
\be
(q_{1},  q_{2}, q_{3}, q_{4}, q_{5}) = \frac{\alpha_{+}}{x-1}\,(1,1,1,1,0)+\cdots = 
\frac{\alpha_{-}}{x+1}\,(1,1,1,1,0)+\cdots.
\ee
Inversion symmetry reads
\be
q_{1}(x) = -q_{2}(1/x), \qquad
q_{3}(x) = 2\,\pi\,m-q_{4}(1/x), \qquad
q_{5}(x) = q_{5}(1/x),
\ee
where $m\in\mathbb Z$ is a winding number. The asymptotic values of the quasi-momenta for a length $L$ state with energy and spin $E$, $S$ are
\be
\label{eq:asym}
\left(\begin{array}{c} q_{1}(x) \\ q_{2}(x) \\ q_{3}(x) \\ q_{4}(x) \\ q_{5}(x) \end{array}\right) = 
\frac{1}{2\,g\,x}\,\left(\begin{array}{l} 
L+E+S \\
L+E-S \\
L-M_{r}+M_{s} \\
L+M_{r}-M_{u}-M_{v} \\
M_{v}-M_{u}
\end{array}\right)+\cdots ,
\ee
where $M_{r,u,v}$ are related to the  $SU(4)$ representation of the state 
\be
[d_{1}, d_{2}, d_{3}] = [L-2M_{u}+M_{r}, M_{u}+M_{v}-2M_{r}+M_{s}, L-2M_{v}+M_{r}].
\ee

\subsection{Semiclassical quantization}

Semiclassical quantization is achieved by perturbing quasi-momenta introducing extra poles that shift the quasi-momenta
$q_{i}\to q_{i}+\delta q_{i}$. Virasoro constraints and inversion properties of the variations $\delta q_{i}$ follow from those of the $q_{i}$'s. In order to find the asymptotic expression of $\delta q_{i}$ in terms of the number $N_{ij}$ of extra fluctuations we can look at the details of polarized states and obtain 
\be
\label{eq:asym2}
\left(\begin{array}{c} \delta q_{1}(x) \\ \delta q_{2}(x) \\ \delta q_{3}(x) \\ \delta q_{4}(x) \\ \delta q_{5}(x) \end{array}\right) = 
\frac{1}{2\,g\,x}\,\left(\begin{array}{ccc} 
\delta E+N_{19}+2\,N_{1, 10} & +N_{15}+N_{16}+N_{17}+N_{18} & \\
\delta E+2\,N_{29}+N_{19} & +N_{25}+N_{26}+N_{27}+N_{28} & \\
& -N_{18}-N_{28} & -N_{35}-N_{36}-N_{37} \\
& -N_{17}-N_{27} & -N_{45}-N_{46}-N_{37} 	\\
& +N_{15}-N_{16}+N_{25}-N_{26} & +N_{35}-N_{36}+N_{45}-N_{46}
\end{array}\right).
\ee
The off-shell frequencies $\Omega^{ij}(x)$ are defined in order to have 
\be
\delta E = \sum_{n, ij} N^{ij}_{n}\,\Omega^{ij}(x^{ij}_{n}),
\ee
where the sum is over all pairs $(ij)\equiv (ji)$ of physical polarizations and integer values of $n$ with 
\be
\label{eq:pole}
q_{i}(x^{ij}_{n})-q_{j}(x^{ij}_{n}) = 2\,\pi\,n.
\ee
Also, the residues at the extra poles are
\be
\delta q_{i}(x) = k_{ij}\,N_{n}^{ij}\,\frac{\alpha(x^{ij}_{n})}{x-x_{n}^{ij}},\quad\mbox{with}\quad
\alpha(x) = \frac{1}{2\,g}\,\frac{x^{2}}{x^{2}-1},
\ee
and $k_{ij}=0, \pm 1, \pm 2$ are the coefficients of $N_{ij}$ in (\ref{eq:asym2}).
%
%
By linear combination of frequencies and inversion (as in the \maldafive case), we can derive all  the off-shell frequencies in terms of two fundamental ones
\be
\Omega_{A}(x) = \Omega^{15}(x), \qquad \Omega_{B}(x) = \Omega^{45}(x).
\ee
Their explicit expressions turns out to be 
\ba
\Omega^{29} &=&  2\,\left[-\Omega_{A}(1/x)+\Omega_{A}(0)\right], \nonumber \\
\Omega^{1, 10} &=&  2\,\Omega_{A}(x),\nonumber \\
\Omega^{19} &=&  \Omega_{A}(x)-\Omega_{A}(1/x)+\Omega_{A}(0), \nonumber \\
\Omega^{37} &=&\Omega_{B}(x)-\Omega_{B}(1/x)+\Omega_{B}(0), \nonumber \\
\Omega^{35}  = \Omega^{36} &=& -\Omega_{B}(1/x)+\Omega_{B}(0),\nonumber \\
\Omega^{45} = \Omega^{46} &=& \Omega_{B}(x), \nonumber \\
\Omega^{17} &=& \Omega_{A}(x)+\Omega_{B}(x), \nonumber \\
\Omega^{18} &=& \Omega_{A}(x)-\Omega_{B}(1/x)+\Omega_{B}(0), \nonumber \\
\Omega^{27} &=& \Omega_{B}(x)-\Omega_{A}(1/x)+\Omega_{A}(0), \nonumber \\
\Omega^{28} &=& -\Omega_{A}(1/x)+\Omega_{A}(0)-\Omega_{B}(1/x)+\Omega_{B}(0), \nonumber \\
\Omega^{15} = \Omega^{16} &=& \Omega_{A}(x), \nonumber \\
\Omega^{25} = \Omega^{26} &=& -\Omega_{A}(1/x)+\Omega_{A}(0).
\ea

\section{The folded string in \adscp}

We present the algebraic curve for the folded string in \adscp\ closely following  the notation of  \cite{Gromov:2008fy}. In terms of the semiclassical variables
\be
\mathcal S = \frac{S}{4\,\pi\,g}, \qquad
\mathcal J = \frac{J}{4\,\pi\,g},
\ee
the energy of the folded string can be expanded according to 
\be
E = 4\,\pi\,g\,\,\mc E_{0}(\mc J, \mc S)+E_{1}(\mc J, \mc S)+\mc O\left(\frac{1}{g}\right),
\ee
where the small $\mc S$ expansion of the classical contribution $\mc E_{0}$ reads
\be
\label{eq:classical}
 \mathcal E_{0}= \mathcal J+\frac{\sqrt{\mathcal J^{2}+1}}{\mathcal J}\,\mathcal S-\frac{\mathcal J^{2}+2}{4\,\mathcal J^{3}(\mathcal J^{2}+1)}\,\mathcal S^{2}+ 
 \frac{3\,\mathcal J^{6}+13\,\mathcal J^{4}+20\,\mathcal J^{2}+8}{16\,\mathcal J^{5}\,(\mathcal J^{2}+1)^{5/2}}\,\mathcal S^{3}+\cdots.
\ee

\subsection{Quasi-momenta}

The quasi momenta are closely related to those of the \ads folded string since motion is still in $AdS_{3}\times S^{1}$
and the $\mathbb{CP}^{3}$ part of the background plays almost no role. The only non trivial case is 
\ba
q_{1}(x) &=& \pi\,f(x)\,\left\{-\frac{J}{4\,\pi\,g}\,\left(\frac{1}{f(1)\,(1-x)}-\frac{1}{f(-1)(1+x)}\right)+ \right. \\
&& \left. -\frac{4}{\pi\,(a+b)(a-x)(a+x)}\left[
(x-a)\,\mathbb K\left(\frac{(b-a)^{2}}{(b+a)^{2}}\right)+ \right.\right. \nonumber \\
&& \left.\left. + 2\,a\,\Pi\left(\left.
\frac{(b-a)(a+x)}{(a+b)(x-a)} \right| \frac{(b-a)^{2}}{(b+a)^{2}}
\right)
\right]
\right\}-\pi.\nonumber
\ea
where the branch points obey $1<a<b$ and 
\be
f(x) = \sqrt{x-a}\,\sqrt{x+a}\,\sqrt{x-b}\,\sqrt{x+b},
\ee
\ba
S &=& 2\,g\,\frac{ab+1}{ab}\,\left[b\,\mathbb E\left(1-\frac{a^{2}}{b^{2}}\right)
-a\,\mathbb K\left(1-\frac{a^{2}}{b^{2}}\right)\right], \nonumber \\
J &=& \frac{4\,g}{b}\,\sqrt{(a^{2}-1)(b^{2}-1)}\,\mathbb K\left(1-\frac{a^{2}}{b^{2}}\right). \\
E &=& 2\,g\,\frac{ab-1}{ab}\,\left[b\,\mathbb E\left(1-\frac{a^{2}}{b^{2}}\right)
+a\,\mathbb K\left(1-\frac{a^{2}}{b^{2}}\right)\right].\nonumber
\ea
The other quasi-momenta are
\ba
&& q_{2}(x) = -q_{1}(1/x), \\
&& q_{3}(x) = q_{4}(x)  = \frac{J}{2\,g}\,\frac{x}{x^{2}-1}.  \\
&& q_{5}(x) = 0.
\ea
The above expressions are valid for a folded string with minimal winding. Adding winding is trivial at the classical level, but requires non trivial changes at the one-loop level (see for instance \cite{Gromov:2011bz}
for a detailed analysis of the \ads\ case).

The independent off-shell frequencies can be determined by the methods of \cite{Gromov:2008ec}. The result is rather simple and reads~\footnote{Notice the important relation
$\Omega_{B}(x) = -\Omega_{B}(1/x)+\Omega_{B}(0)$.
}
\ba
&& \Omega_{A}(x) =  \frac{1}{ab-1}\left(1-\frac{f(x)}{x^{2}-1}\right),   \\
&& \Omega_{B}(x) = \frac{\sqrt{a^{2}-1}\,\sqrt{b^{2}-1}}{ab-1}\,\frac{1}{x^{2}-1}.
\ea

\section{Integral representation for the one-loop correction to the energy}

The one-loop shift of the energy is given in full generality by the following sum of zero point energies
\be
\label{eq:one-loop-correction}
E_{1} = \frac{1}{2}\,\sum_{n=-\infty}^{\infty}\,\sum_{ij}(-1)^{F_{ij}}\,\omega^{ij}_{n},\qquad
\omega_{n}^{ij} = \Omega^{ij}(x^{ij}_{n}),
\ee
where the sum over $ij$ is over the $8_{B}+8_{F}$ physical polarizations and $x^{ij}_{n}$ is the unique solution 
to the equation (\ref{eq:pole})  under the condition $|x_{n}^{ij}|>1$~\footnote{ If it happens that for some $ij$ and $n$ the above equation has no solution, then we shall say that the polarization $(ij)$ has the { missing mode} $n$. Missing modes can be treated according to the procedure
discussed in \cite{Gromov:2008ec}.
}.

In the same spirit as \cite{Gromov:2011de,Gromov:2011bz}, the infinite sum over on-shell frequencies can be 
evaluated by contour integration in the complex plane. The result is quite similar to the \ads\ one and 
reads
\be
E_{1} = E_{1}^{\rm anomaly, 1}+E_{1}^{\rm anomaly, 2}+E_{1}^{\rm dressing}+E_{1}^{\rm wrapping},
\ee
with~\footnote{Here, $x(z) = z+\sqrt{z^{2}-1}$. Also the anomaly contributions are computed integrating on the upper half complex plane.}
\ba
E_{1}^{\rm anomaly, 1} &=& 2\,\int_{a}^{b}\,\frac{dx}{2\,\pi\,i}\left[\Omega^{1,10}(x)-\Omega^{1,10}(a)\right]\,
\partial_{x}\,\log\sin q_{1}(x), \\
E_{1}^{\rm anomaly, 2} &=& -2\times 2\,\int_{a}^{b}\,\frac{dx}{2\,\pi\,i}\left[\Omega^{1,5}(x)-\Omega^{1,5}(a)\right]\,\partial_{x}\,\log\sin \frac{q_{1}(x)}{2}, \\
E_{1}^{\rm dressing} &=& \sum_{ij}(-1)^{F_{ij}}\,\int_{-1}^{1}\frac{dz}{2\,\pi\,i}
\,\Omega^{ij}(z)\,\partial_{z}\frac{i\,\left[q_{i}(z)-q_{j}(z)\right]}{2},\\
E_{1}^{\rm wrapping} &=& \sum_{ij}(-1)^{F_{ij}}\,\int_{-1}^{1}\frac{dz}{2\,\pi\,i}
\,\Omega^{ij}(z)\,\partial_{z}\log(1-e^{-i\,(q_{i}(z)-q_{j}(z))}),
\ea 
As in \ads, the labeling of the various contributions reminds their physical origin. In particular, dressing and wrapping contributions have been separated in order to split the asymptotic contribution from finite size effects. As in \ads, the anomaly terms are special 
contributions arising from the deformation of contours and ultimately due to the presence of the algebraic curve cuts.
The representation (\ref{eq:one-loop-correction}) is a compact formula for $E_{1}$ and can be evaluated numerically with minor effort. In order to understand it better, we shall now analyze the short and long string limit. In the former case, 
we shall evaluate the explicit sum over frequencies clarifying the relation with the contour integrals. In the latter, we shall extract the analytical expansion at large spin directly from (\ref{eq:one-loop-correction}).

\section{Short string limit}

The short string limit is generically $\mc S\to 0$. Regarding $\mc J$, we shall consider two cases. The first amounts to 
keeping $\mc J$ fixed, expanding in the end each coefficient of powers of $\mc S$ at small $\mc J$. This is 
precisely the procedure worked out in \cite{Gromov:2011bz} in \ads. In the second case, we shall keep the ratio
$\rho = \mc J/\sqrt\mc S$ fixed as in \cite{Beccaria:2012tu}. The two expansions are related, but not equivalent and provide useful different information.

\subsection{Fixed $\mc J$ expansion}

After a straightforward computation, our main result is 
\ba
\label{eq:GV}
E_{1} &=& 
\bigg(
-\frac{1}{2 \mathcal{J}^2}+\frac{\log (2)-\frac{1}{2}}{\mathcal{J}}+\frac{1}{4}+\mathcal{J} \left(-\frac{3 \,\zeta (3)}{8}+\frac{1}{2}-\frac{\log (2)}{2}\right)-\frac{3
   \mathcal{J}^2}{16}+\\
   &&+\mathcal{J}^3 \left(\frac{3 \,\zeta (3)}{16}+\frac{45 \,\zeta (5)}{128}-\frac{1}{2}+\frac{3 \log (2)}{8}\right)+
   \cdots
\bigg)\,\mc S+ \nonumber \\
&& +\bigg(
\frac{3}{4 \mathcal{J}^4}+\frac{\frac{1}{2}-\log (2)}{\mathcal{J}^3}-\frac{1}{8 \mathcal{J}^2}+\frac{\frac{1}{16}-\frac{3 \,\zeta (3)}{4}}{\mathcal{J}}-\frac{1}{8}+\mathcal{J} \left(\frac{69 \,\zeta
   (3)}{64}+\frac{165 \,\zeta (5)}{128}-\frac{27}{32}+\frac{\log (2)}{2}\right)+\nonumber \\
   && +\frac{3 \mathcal{J}^2}{8}+\mathcal{J}^3 \left(-\frac{163 \,\zeta (3)}{128}-\frac{405 \,\zeta (5)}{256}-\frac{875 \,\zeta
   (7)}{512}+\frac{235}{128}-\log (2)\right)+\cdots
\bigg)\,\mc S^{2}+ \nonumber \\
&& + 
\bigg(
-\frac{5}{4 \mathcal{J}^6}+\frac{\frac{3 \log (2)}{2}-\frac{3}{4}}{\mathcal{J}^5}+\frac{\frac{9 \,\zeta (3)}{16}+\frac{1}{16}}{\mathcal{J}^3}+\frac{1}{16 \mathcal{J}^2}+\frac{\frac{45 \,\zeta
   (3)}{64}+\frac{75 \,\zeta (5)}{256}-\frac{7}{32}+\frac{\log (2)}{8}}{\mathcal{J}}+\frac{11}{64}+\nonumber \\
   &&  +\mathcal{J} \left(-\frac{89 \,\zeta (3)}{32}-\frac{745 \,\zeta (5)}{256}-\frac{3815 \,\zeta
   (7)}{2048}+2-\frac{33 \log (2)}{32}\right)-\frac{465 \mathcal{J}^2}{512}+\nonumber \\
   && + \mathcal{J}^3 \left(\frac{5833 \,\zeta (3)}{1024}+\frac{1585 \,\zeta (5)}{256}+\frac{98035 \,\zeta (7)}{16384}+\frac{259455
   \,\zeta (9)}{65536}-\frac{405}{64}+\frac{775 \log (2)}{256}\right)+\cdots
\bigg)\,\mc S^{3}+ \cdots\nonumber
\ea
This expansion is rather similar to the one derived in \cite{Gromov:2011bz} for \ads, but there are two remarkable differences:
\begin{enumerate}
\item The leading terms at small $\mc J$ are $\mc O(\mc S^{n}/\mc J^{2n})$. Instead, they were $\mc O(\mc S^{n}/\mc J^{2n-1})$ in \ads. Also, there are terms with all parities in $\mc J$ while in \ads, there appear only terms odd under $\mc J\to -\mc J$. The additional terms are important and we shall discuss them in more details later. Remarkably, they 
imply that if one scales $\mc J\sim \sqrt\mc S$ they give a constant contribution in the short string limit. This is different 
from \ads\ where the energy correction vanishes like $\sqrt\mc S$ in this regime.

\item There are terms proportional to $\log(2)$. As we  discuss in App.~(\ref{app:log2}), 
these terms can be removed by expressing the energy correction in terms of the coupling in the 
world-sheet scheme. The scheme dependence is universal and agrees with that found in 
\cite{McLoughlin:2008he} for a circular string solution and in \cite{Abbott:2010yb}
for the giant magnon.

\end{enumerate}

\subsection{Fixed $\rho = \mc J / \sqrt\mc S$ expansion}

The result in this limit is 
\ba
\label{eq:our-expansion}
&& E_{1} = -\frac{1}{2}\,\mc C(\rho, \mc S)+a_{01}(\rho)\,\sqrt\mathcal S+ a_{1,1}(\rho)\,\mathcal S^{3/2}+\mc O(\mc S^{5/2}),
\ea
where
\ba
a_{1,0}(\rho) &=& \frac{2\,\log (2)-1}{2\,\sqrt{\rho^{2}+2}}, \\
a_{1,1}(\rho) &=& -\frac{\log (2)\left(2 \rho ^4+6 \rho ^2+3\right)}{4 \left(\rho ^2+2\right)^{3/2}}+\frac{8 \rho ^4+25 \rho ^2+16}{16 \left(\rho ^2+2\right)^{3/2}}-\frac{3 \left(\rho
   ^2+3\right) \zeta (3)}{8 \sqrt{\rho ^2+2}},
\ea
and $\mc C$ is related to the branch cut endpoints by the formula
\be
\label{eq:theC}
\mc C = \frac{\sqrt{(a^{2}-1)\,(b^{2}-1)}}{1-a\,b}+1.
\ee
Its expansion at small $\mc S$ with fixed $\rho=\mc J/\sqrt\mc S$ is 
\be
\mc C = 1-\frac{\rho}{\sqrt{\rho^{2}+2}}-\frac{2\,\rho^{3}+5\,\rho}{4\,(\rho^{2}+2)^{3/2}}\,\mc S + 
\frac{\rho\,(12\,\rho^{6}+68\,\rho^{4}+126\,\rho^{2}+73)}{32\,(\rho^{2}+2)^{5/2}}\,\mc S^{2}+\cdots
\ee
Expanding $E_{1}$ at large $\rho$ we  partially resum the calculation at fixed $\mathcal J$. Just to give an 
example, from the expansion 
\be
-\frac{1}{2}\left(1-\frac{\rho}{\sqrt{\rho^{2}+2}}\right) = -\frac{1}{2 \rho ^2}+\frac{3}{4 \rho ^4}-\frac{5}{4 \rho ^6}+\frac{35}{16 \rho
   ^8}-\frac{63}{16 \rho ^{10}}+\cdots,
\ee
we read the coefficients of {\bf all } terms $\sim \mc S^{n}/\mc J^{2n}$. The first ones are of course in agreement
with (\ref{eq:GV}). As another non trivial example, the large $\rho$ expansion of $a_{11}(\rho)$ is 
\ba
a_{11}(\rho) &=& \rho  \left(-\frac{3 \zeta
   (3)}{8}+\frac{1}{2}-\frac{\log
   (2)}{2}\right)+\frac{\frac{1}{16}-\frac{3 \zeta
   (3)}{4}}{\rho }+\frac{\frac{9 \zeta
   (3)}{16}+\frac{1}{16}}{\rho ^3}+\\
   &&+\frac{-\frac{3
   \zeta (3)}{4}-\frac{1}{32}-\frac{\log
   (2)}{4}}{\rho ^5}+\frac{\frac{75 \zeta
   (3)}{64}-\frac{5}{32}+\frac{15 \log (2)}{16}}{\rho
   ^7}+\cdots, \nonumber
\ea
and allows to read the coefficients of all terms $\sim \mc S^{n} / \mc J^{2n-3}$.

\subsection{Summation issues}

The explicit sum over the infinite number of on-shell frequencies requires some care and a definite prescription 
since the sums are not absolutely convergent due to physically sensible cancellations between bosonic and fermionic contributions.
As discussed in \cite{Gromov:2008fy}, the following summation prescription is natural from the point of view
of the algebraic curve (see \cite{Bandres:2009kw} for a different prescription)~\footnote{Notice that we
exploit the $x\to -x$ symmetry of the classical algebraic curve as well as triviality of zero mode corrections.}
\be
E_{1} = \sum_{n=1}^{\infty} K_{n}, 
\ee
where $K_{n}$ is a particular grouping of heavy and light modes
\be
\label{eq:Kdef}
K_{n} = \left\{\begin{array}{cc}
\omega^{\rm heavy}_{n}+\omega^{\rm light}_{n/2} & \quad n\in 2\,\mathbb Z\\ \\
\omega^{\rm heavy}_{n} & \quad n\not\in 2\,\mathbb Z,
\end{array}\right.
\ee
with
\ba
\omega_{n}^{\rm heavy} &=& \omega^{(AdS, 1)}_{n}+\omega^{(AdS, 2)}_{n}+\omega^{(AdS, 3)}_{n}+
\omega^{(\mathbb{CP}, 1)}_{n}-2\,\omega^{(F, 1)}_{n}-2\,\omega^{(F, 2)}_{n}, \\
\omega_{n}^{\rm light} &=& 4\,\omega^{(\mathbb{CP}, 2)}_{n}-2\,\omega^{(F, 3)}_{n}-2\,\omega^{(F, 4)}_{n}.
\ea
The short string expansion of $K_{n}$ takes the form
\ba
K_{p} &=& (-1)^{p}\,\mc C+\widehat K_{p}
\ea
where $\mc C$, given in (\ref{eq:theC}),  is independent on $p$ and the sum of $\widehat K_{p}$ (which start 
at $\mc O(\mc S)$) is convergent.
The alternating constant $\mc C$ poses some problems because we have to give a meaning to 
\be
-\mc C+\mc C-\mc C+\mc C+\cdots.
\ee
An analysis of the integral representation shows that it automatically selects the choice
\be
\label{eq:alternating}
-\mc C+\mc C-\mc C+\mc C+\cdots \equiv -\frac{1}{2}\,\mc C
\ee
Later, we shall provide various consistency checks of this prescription. In particular, we shall see that  
it is necessary in order to match the
asymptotic Bethe Ansatz equations when wrapping effects are subtracted.
Notice also that the expansion of $\mc C$ at fixed $\mc J$ is 
\be
\mc C = \frac{\mathcal{S}}{\mathcal{J}^2 \sqrt{\mathcal{J}^2+1}}-\frac{\left(3 \mathcal{J}^4+11 \mathcal{J}^2+6\right) \mathcal{S}^2}{4 \mathcal{J}^4
   \left(\mathcal{J}^2+1\right)^2}+\frac{12\,\mc J^{8}+75\,\mc J^{6}+173\,\mc J^{4}+140\,\mc J^{2}+40}{16\,\mc J^{6}\,(\mc J^{2}+1)^{7/2}}\,\mc S^{3}+\cdots,
 \ee
 so, upon expanding at small $\mc J$, it provides precisely 
 the terms with even/odd $\mc J$ exponents in the coefficients of the odd/even powers of $\mc S$ in (\ref{eq:GV}).

Apart from the $\mc C$ term, the integral representation implements the Gromov-Mikhailov (GM) prescription. The reason is that the singularities at $|x|=1$ are avoided by implicitly encircling them by a small circumference. This cut-off on $|x-1|$ translates in a bound on the highest mode $n$ that correlates heavy/light polarizations according to GM. In other words the highest mode for light polarizations is asymptotically half the highest mode for heavy polarizations. 

As a numerical check of the agreement between the integral representation and the series expansion, 
we fix $\rho=1$ in table (\ref{tab:check1})  and show
the value of $E_{1}$ from our analytical resummation and result from the integral. The agreement is very good already 
at moderately small $\mc S$.

\begin{table}[htb]
\begin{center}
\begin{tabular}{c|ll}
$\mc S$ & $E_{1}$ from (\ref{eq:our-expansion}) & $E_{1}$ \\
\hline
1/10 & -0.18790 & -0.17987  \\
1/50 & -0.19461 & -0.19443 \\
1/100 & -0.19934 & -0.19930 \\
1/300 & -0.20449 & -0.20448 \\
1/500 & -0.206075 & -0.206075
\end{tabular}
\caption{Comparison between resummation at fixed ratio
$\rho=1$ and integral representation. The asymptotic value for $\mc S\to 0$ is $(\sqrt 3-3)/6\simeq -0.211$, but already at $\mc S = 1/500$ we have 6 digits agreement.
}
\label{tab:check1}
\end{center}
\end{table}

\begin{figure}[htb]
\begin{center}
\includegraphics[width=10cm]{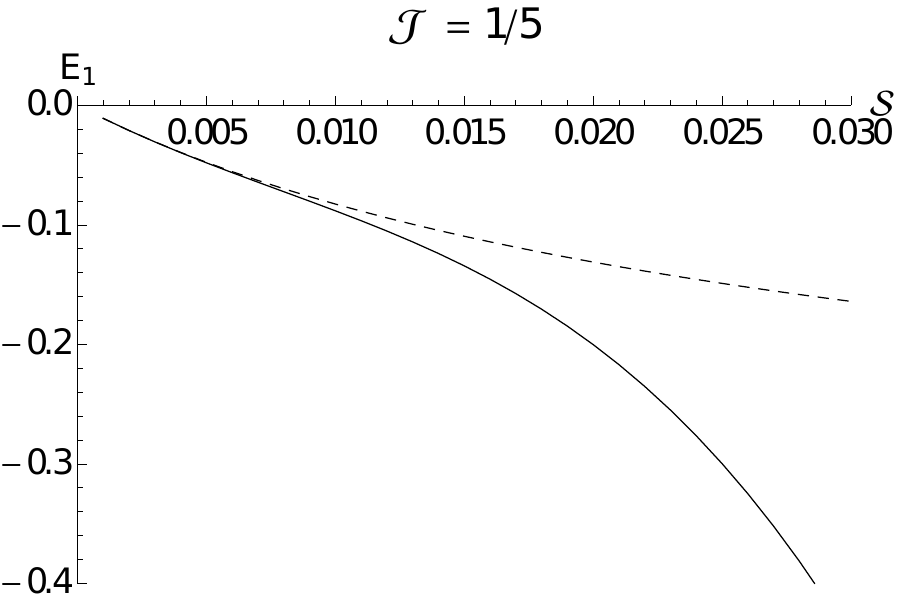}
\caption{Asymptotic expansion (\ref{eq:GV}) [solid line] 
and  exact numerical $E_{1}$ [dashed line] as functions of $\mc S$ at $\mc J=1/5$.}
\label{fig:check}
\end{center}
\end{figure}

\noindent
A similar check at fixed $\mc J$ is shown in Fig.~(\ref{fig:check}) where we plot the asymptotic expansion (\ref{eq:GV}) and the exact numerical $E_{1}$ as functions of $\mc S$ at $\mc J=1/5$.

\section{The slope function}
\label{sec:slope}

The one-loop correction $E_{1}$ tends to zero linearly with $\mc S$ when $\mc S\to 0$ at fixed $\mc J$.
The slope ratio
\be
\sigma(\mc J) = \lim_{\mc S\to 0}\frac{E_{1}(\mc S, \mc J)}{\mc S},
\ee
is an important quantity  related to the conjectures in \cite{Basso:2011rs}~\footnote{The exact slope mentioned in 
\cite{Basso:2011rs} is the coefficient of $S$ in the expansion of $E^{2}$. This line of analysis is suggested by the simplicity of the marginality condition in \ads (see \cite{Tseytlin:2003ac} for a general discussion). Here, it is simpler to discuss the quantity $\sigma(\mc J)$.} . It is known that it does not receive dressing corrections both in \ads\ and in \adscp\ since such contributions start at 
order $\mc S^{2}$ \cite{Basso:2011rs}. It also 
does not receive wrapping corrections in $AdS_{5}$.  Instead, in the case of $AdS_{4}$ the slope has a non vanishing wrapping contribution. For instance, a rough evaluation at $\mc J=1$ gives  a definitely non zero value around $-0.042$.

\bigskip
Indeed, an analytical calculation shows that  the wrapping contribution to the slope in \adscp\ is exactly
\be
\sigma^{\rm wrap}(\mc J) = \sum_{n=-\infty}^{\infty}\sigma_{n} = -\frac{1}{2\,\mc J}\,\sum_{n=-\infty}^{\infty}\frac{(-1)^{n}}{
\sqrt{\mc J^{4}+(n^{2}+1)\,\mc J^{2}+n^{2}}}.
\ee
This formula is in perfect agreement with numerics since for instance
\be
\sigma^{\rm wrap}(\mc J=1) = -0.041777654879558824814\dots.
\ee
The large $\mc J$ limit of this expression is exponentially suppressed as it should
\be
\sigma^{\rm wrap}(\mc J)  = -\frac{\sqrt 2}{\mc J^{5/2}}\,e^{-\pi\,\mc J}+\cdots
\ee
To analyze the small $\mc J$ limit it is convenient to split this contribution into the $n=0$ term plus the rest. The result is very intriguing. For the $n=0$ term, 
we find 
\ba
\sigma^{\rm wrap}_{n=0} = -\frac{1}{2\,\mc J^{2}\,\sqrt{\mc J^{2}+1}} = 
-\frac{1}{2 \mathcal{J}^2}+\frac{1}{4}-\frac{3 \mathcal{J}^2}{16}+\frac{5 \mathcal{J}^4}{32}-\frac{35 \mathcal{J}^6}{256}+\frac{63 \mathcal{J}^8}{512}+\cdots\,.
\ea
This is precisely the set of terms even under $\mc J\to -\mc J$ in the full slope which is the first term of  (\ref{eq:GV}).
Similarly, we can consider the rest of $\sigma^{\rm wrap}$ and expand at small $\mc J$. We find 
\ba
\sum_{n\neq 0}\sigma^{\rm wrap}_n &=& \frac{\log (2)}{\mathcal{J}}+\mathcal{J} \left(-\frac{3 \zeta(3)}{8}-\frac{\log (2)}{2}\right)+\mathcal{J}^3 \left(\frac{3 \zeta(3)}{16}+\frac{45 \zeta(5)}{128}+\frac{3 \log (2)}{8}\right)+\nonumber \\
&&+\mathcal{J}^5
   \left(-\frac{9 \zeta(3)}{64}-\frac{45 \zeta(5)}{256}-\frac{315 \zeta(7)}{1024}-\frac{5 \log (2)}{16}\right)+O\left(\mathcal{J}^6\right).
\ea
Comparing again with (\ref{eq:GV}), we see that we are reproducing all the irrational terms of the slope, involving zeta functions or $\log(2)$. The remaining terms 
are the same as in \ads~\footnote{
This is due to the fact that the BAE are essentially the same as for $\mathfrak{sl}(2)$ sector in \ads. 
This is however a nontrivial test that all is done correctly.
},
\be
\sigma(\mc J) -\sigma^{\rm wrap}(\mc J) = -\frac{1}{2\,\mc J}+\frac{\mc J}{2}-\frac{\mc J^{3}}{2}+\cdots.
\ee
Thus, we are led to the following expression for the one-loop full slope 
\be
\sigma(\mc J) = -\frac{1}{2\,\mc J}\left[\frac{1}{\mc J^{2}+1}+\sum_{n=-\infty}^{\infty}\frac{(-1)^{n}}{
\sqrt{\mc J^{4}+(n^{2}+1)\,\mc J^{2}+n^{2}}}
\right].
\ee

The above analysis of the slope is a confirmation  that the various terms in (\ref{eq:GV}) are organized in the expected way. The asymptotic contribution is precisely the same as in \ads, while wrapping is different and is exponentially
suppressed for large operators. This is a property of the integral representation and a confirm that the 
prescription (\ref{eq:alternating}) is correct.

\subsection{Weak coupling}

It is interesting to evaluate the slope at weak coupling.
 In principle, this  requires the knowledge of the anomalous dimensions of 
short $\mathfrak{sl}(2)$ operators in closed form as a function of the spin at a certain length ({\em i.e.} twist, in the gauge theory
dictionary). This information is 
available for the asymptotic contribution, but not for the wrapping, which is only known as a series expansion 
at large spin and low twist \cite{Beccaria:2009ny,Beccaria:2010kd}.
 Nevertheless, if we are interested in the correction to the slope only (so, just the first term at small spin), then the L\"uscher form of the wrapping correction
presented in \cite{Beccaria:2009ny} is enough~\footnote{We kindly thank B. Basso for this important remark.}. 
At twist-1, and following the notation of \cite{Beccaria:2009ny}, the wrapping correction enters at four loops
and is expressed by the following function of the integer spin $N$ of the gauge theory operator
\be
\gamma_{4}^{\rm wrapping}(N) = \gamma_{2}(N)\,\mc W(N),
\qquad \gamma_{2}(N) = 4\,[S_{1}(N)-S_{-1}(N)].
\ee
Here, $S_{a}(N)$ are generalized harmonic sums while $\mc W(N)$ is a complicated expression depending on the 
Baxter polynomial $Q_{N}(u)$ associated with the Bethe roots. The first factor $\gamma_{2}(N)$ 
is nothing but the two-loop
anomalous dimension of the twist-1 operators. In the small $N$ limit, it starts at $\mc O(N)$. Thus, the 
factor $\mc W(N)$ can be evaluated at $N=0$ where the Baxter polynomial trivializes $Q_{0}(u)=1$.
After a straightforward calculation, one finds that (on the even $N$ branch),
\be
\gamma_{4}^{\rm wrapping}(N) = -\frac{\pi^{4}}{3}\,N+\mc O(N^{2}).
\ee
So, even at weak coupling, we find a correction to the slope coming from the wrapping 
terms~\footnote{Notice that 
the reason why such a contribution is absent in \ads\ is simply that the factor analogous to $\gamma_{2}(N)$
is squared in the wrapping contribution. This leads immediately to a contribution to the slope of order $\mc O(N^{2})$.}.

\section{Long string limit}

The large $S$ behaviour of the one-loop energy $E_{1}$ can be computed starting from the integral representation.
Let us first summarize the result valid for \ads\ from \cite{Gromov:2011de}. We scale $\mc J$ with $\mc S$ for $\mc S\gg 1$ according to 
\be
\mc J = \frac{\ell}{\pi}\,\log\left(\frac{8\,\pi\,\mc S}{\sqrt{\ell^{2}+1}}\right),
\ee
where we assume $\ell>0$~\footnote{This means that the case $\ell=0$, or $\mc J=0$ has to be treated
separately as discussed in \cite{Gromov:2011de}.}. Then, the one-loop energy correction can be written
\be
E_{1}^{AdS_{5}} = f_{10}^{AdS_{5}} (\ell)\,\log\left(\frac{8\,\pi\,\mc S}{\sqrt{\ell^{2}+1}}\right)+f_{11}^{AdS_{5}} (\ell)+\frac{c^{AdS_{5}}}{\log\left(\frac{8\,\pi\,\mc S}{\sqrt{\ell^{2}+1}}\right)}+\cdots ,
\ee
with
\ba
f_{10}^{AdS_{5}}(\ell) &=& \frac{\sqrt{\ell^2+1}+2 \left(\ell^2+1\right) \log
   \left(\frac{1}{\ell^2}+1\right)-\left(\ell^2+2\right)
   \log
   \left(\frac{\sqrt{\ell^2+2}}{\sqrt{\ell^2+1}-1}\right)-1
   }{\pi  \sqrt{\ell^2+1}}, \\
f_{11}^{AdS_{5}}(\ell) &=& \frac{2 \left(\log
   \left(1-\frac{1}{\left(\ell^2+1\right)^2}\right)+2
   \sqrt{\ell^2+1} \cot ^{-1}\left(\sqrt{\ell^2+1}\right)+2
   \coth ^{-1}\left(\sqrt{\ell^2+1}\right)-2 \ell \cot
   ^{-1}(\ell)\right)}{\pi  \sqrt{\ell^2+1}}.\nonumber, \\
c^{AdS_{5}}(\ell) &=& -\frac{\pi}{12\,(\ell^{2}+1)}.\nonumber
\ea

The expansion in \adscp\ can be derived in the same way as in \cite{Gromov:2011de} and the result is simply
\be
\label{eq:AdS4-largeS}
E_{1}^{AdS_{4}} =f_{10}^{AdS_{4}} (\ell)\,\log\left(\frac{8\,\pi\,\mc S}{\sqrt{\ell^{2}+1}}\right)+f_{11}^{AdS_{4}} (\ell)+\frac{c^{\rm AdS_{4}}}{\log\left(\frac{8\,\pi\,\mc S}{\sqrt{\ell^{2}+1}}\right)}+\cdots ,
\ee
with 
\ba
f_{10}^{AdS_{4}}(\ell) &=& \frac{1}{2}\,f_{10}^{AdS_{5}}(\ell), \nonumber \\
f_{11}^{AdS_{4}}(\ell) &=& \frac{1}{2}\,f_{11}^{AdS_{5}}(\ell), \\
c^{AdS_{4}}(\ell) &=& 2\,c^{AdS_{5}}(\ell) = -\frac{\pi}{6\,(\ell^{2}+1)}.\nonumber
\ea
This formula can be easily checked numerically from the explicit evaluation of the integral representation.
Notice that the simple $\frac{1}{2}$ rule for the leading two terms is in agreement with the result of \cite{Beccaria:2009wb}. The correction $\sim 1/\log \mc S$ comes from the anomaly terms. It is twice bigger than in SYM. 

The explanation of this fact is as follows~\footnote{We thank B. Basso for clarifying this point as well as the $\ell\to 0$ limit.}.
The low energy effective theory of the Gubser-Klebanov-Polyakov (GKP) 
string in \adscp\  has two massless modes at finite chemical potential $\ell$. Namely, 
one massless Dirac Fermion and one massless boson that gives a central charge 2 (in \ads\ 
one has only one massless boson giving central charge 1). Also, concerning the $\ell\to 0$ limit, 
the other low-energy modes acquire a mass proportional to $\ell$ at small $\ell$ and their
 contribution is exponentially suppressed with the effective length $\log \mc S$ at fixed $\ell$. When $\ell\to 0$
  they become massless and contribute at leading order to 5 units of central charge 
  (there are actually 4 bosons with mass $\ell$ and one with mass $\ell/2$ while there were only four with mass $\ell$ in \ads\ ). In other words it should be true that in the small $\ell$ limit the $1/\log\mc S$  gets corrected by 
  5 extra units of central charge giving a total $-(2+5)\frac{\pi}{12\,\log\mc S}$
for the energy of the vacuum state (i.e. the twist 1 state of the theory). Indeed,  $2+5=7$ is the correct central charge of the low-energy effective theory on the GKP background \cite{Alday:2008ut,Alday:2009zz}. Instead, 
in \ads\  the final result for $\ell\to 0$ (i.e. for twist 2) was coming with $1+4 = 5$ 
units of central charge, which is the correct central charge of the  $O(6)$ model.

\section{Relation with marginality condition}

Let us define $\Lambda\equiv\lambda$ in \ads, and $\Lambda = 16\,\pi^{2}\,g^{2}$ in \adscp. The role of $\Lambda$
is to  emphasize the close analogy between the expressions in the two cases.
For the folded string in \ads, the energy admits the following expansion 
\ba
\label{eq:marginality} 
E^{2} &=& J^{2}+\left(A_{1}\,\sqrt\Lambda+A_{2}+\frac{A_{3}}{\sqrt\Lambda}
+\cdots\right)\,S + \left(B_{1}+\frac{B_{2}}{\sqrt\Lambda}+ \frac{B_{3}}{\Lambda}+\cdots\right)\,S^{2}+ \\
&&+
\left(\frac{C_{1}}{\sqrt\Lambda}+\frac{C_{2}}{\Lambda}+\frac{C_{3}}{\Lambda^{3/2}}
+\cdots\right)\,S^{3}+\cdots , \nonumber
\ea
where the following exact formula  for the constants $A_{i}$ has been conjectured in \cite{Basso:2011rs}:
\be
A_{1}\,\sqrt\Lambda+A_{2}+\frac{A_{3}}{\sqrt\Lambda}+\cdots = 2\,\sqrt\Lambda\,Y_{J}(\sqrt\Lambda), \qquad
Y_{J}(x) = \frac{d}{dx}\,\log I_{J}(x).
\ee
Expanding at large $\lambda$, we find the first values
\be
\begin{array}{ccl}
A_{1} &=& 2, \\
A_{2} &=& -1, \\
A_{3} &=& J^{2}-\frac{1}{4}, \\
A_{4} &=& J^{2}-\frac{1}{4}, 
\end{array}\qquad
\begin{array}{ccl}
A_{5} &=& -\frac{1}{4}\,J^{4}+\frac{13}{8}\,J^{2}-\frac{25}{64}, \\
A_{6} &=& -J^{4}+\frac{7}{2}\,J^{2}-\frac{13}{16}, \\
A_{7} &=& \frac{J^6}{8}-\frac{115 J^4}{32}+\frac{1187 J^2}{128}-\frac{1073}{512}.
\end{array}
\ee
Also, it is  know that $B_{1} = \frac{3}{2}$ and $B_{2} = \frac{3}{8}-3\,\zeta(3)$  \cite{Gromov:2011bz} .

\bigskip
The expansion (\ref{eq:marginality}) is very convenient since all powers of $S$ have a coefficient with an expansion 
at large $\Lambda$ starting with a more and more suppressed term. The simplicity of (\ref{eq:marginality}) is a special
feature of the folded string with two cusps. If winding is allowed, it is known that such structure is lost
as discussed in \cite{Gromov:2011bz} (see also the results of \cite{Beccaria:2012tu}).

\bigskip
For the folded string in \adscp, the expansion with fixed $\mc J$~\footnote{Actually when we speak about fixed $\mc J$ we mean small $\mc S$ followed by small $\mc J$.} has the general form (see the Appendices of \cite{Gromov:2011bz})
\be
E = \sqrt\Lambda \,\,\,\mc E_{0} + \sum_{\ell=0}^{\infty}\frac{1}{(\sqrt\Lambda)^{\ell}}\,
\sum_{p=1}^{\infty}\sum_{q=-2p}^{\infty} v_{pq}^{(\ell)}\,\mc J^{q}\,\mc S^{p},
\ee
where the classical energy is~\footnote{Note that there is a typo in the $\mc S^{3}$ term in the introduction 
to \cite{Beccaria:2012tu}}
 \ba
&& \mathcal E_{0}= \mathcal J+\frac{\sqrt{\mathcal J^{2}+1}}{\mathcal J}\,\mathcal S-\frac{\mathcal J^{2}+2}{4\,\mathcal J^{3}(\mathcal J^{2}+1)}\,\mathcal S^{2}+ 
 \frac{3\,\mathcal J^{6}+13\,\mathcal J^{4}+20\,\mathcal J^{2}+8}{16\,\mathcal J^{5}\,(\mathcal J^{2}+1)^{5/2}}\,\mathcal S^{3}+\cdots.
\ea
and the semiclassical computation provides $v^{(0)}_{pq}$ according to the results in (\ref{eq:GV}).

\bigskip
Expanding $E^{2}$, we find that  (\ref{eq:marginality}) takes  the following form 
\ba
\lefteqn{E^{2} - J^{2} = } && \nonumber \\
&&+\bigg[
\left(2-\frac{1}{J}\right) \sqrt{\Lambda }+\left(\frac{2 v^{\text{(1)}}_{1,-2}}{J}-1+2 \log
   (2)\right)+\sqrt{\frac{1}{\Lambda }} \left(
   \frac{2 v^{(2)}_{1,-2}}{J}+
   2 v^{\text{(1)}}_{1,-1}+J^2+\frac{J}{2}\right)+
   \cdots
\bigg]\,S+\nonumber \\
&& + \bigg[
\left(\frac{1}{4 J^4}+\frac{1}{2 J^3}\right) \Lambda +\sqrt{\Lambda }
   \left(-\frac{v^{\text{(1)}}_{1,-2}}{J^4}+\frac{2 v^{\text{(1)}}_{1,-2}}{J^3}+\frac{2
   v^{\text{(1)}}_{2,-4}}{J^3}+\frac{1}{2 J^3}-\frac{\log
   (2)}{J^3}\right)+\cdots
\bigg]\,S^{2}+\nonumber \\
&& + \bigg[
\left(-\frac{3}{4 J^6}-\frac{1}{2 J^5}\right) \Lambda ^{3/2}+\cdots
\bigg]\,S^{3}+\cdots . 
\ea
This structure is different from (\ref{eq:marginality}) since higher powers of $S$ are not associated with terms that are 
more and more suppressed at large $\Lambda$. This is possible since the new terms not present in (\ref{eq:marginality})
are associated with suitable inverse powers of $J$. The same phenomenon  is discussed in  \cite{Gromov:2011bz}
for the folded string in \ads\ with non-trivial  winding. As we discussed in Sec.~(\ref{sec:slope}), wrapping corrections are responsible for these terms.

\subsection{Prediction for short states}

We can provide a prediction for the strong coupling expansion of the energy of short states that in principle
could be tested by TBA calculations. To this aim, we 
can start from  our results at fixed $\rho = \mc J/\sqrt\mc S$, and 
re-expand at large $\Lambda$ the sum of the (scaled) classical energy 
\ba
\label{eq:classical-rho}
\mathcal E_{0} &=&\sqrt{(\rho^{2}+2)\,\mathcal S}\,\bigg[
1+\frac{2\,\rho^{2}+3}{4\,(\rho^{2}+2)}\,\mathcal S-
\frac{4\,\rho^{6}+20\,\rho^{4}+34\,\rho^{2}+21}{32\,(\rho^{2}+2)^{2}}\,\mathcal S^{2}+\cdots
\bigg]
\ea
and the one-loop contribution (\ref{eq:our-expansion}). The result is 
\be
\label{eq:short}
E = (4\,\pi\,g)^{1/2}\,\sqrt{2\,S}-\frac{1}{2}+\frac{\sqrt{2\,S}}{(4\,\pi\,g)^{1/2}}\,\left(
\frac{J\,(J+1)}{4\,S}+\frac{3\,S}{8}-\frac{1}{4}+\frac{1}{2}\,\log(2)\right)+\cdots.
\ee
The same expansion 
where we remark that one of the effect of the $\mc C$ term 
is the constant $\mc O(\widetilde \Lambda^{0})$ contribution.

The same expansion can be written in terms of the coupling $g_{\rm WS}$ in the world-sheet regularization
whose relation with $g$ is \cite{McLoughlin:2008he,Abbott:2010yb}
\be
g = g_{\rm WS}-\frac{\log(2)}{4\pi}+\cdots.
\ee
After this replacement, eq.(\ref{eq:short}) reads  
\be
\label{eq:short}
E = (4\,\pi\,g_{\rm WS})^{1/2}\,\sqrt{2\,S}-\frac{1}{2}+\frac{\sqrt{2\,S}}{(4\,\pi\,g_{\rm WS})^{1/2}}\,\left(
\frac{J\,(J+1)}{4\,S}+\frac{3\,S}{8}-\frac{1}{4}\right)+\cdots,
\ee
without $\log(2)$ term. This is correct since in world-sheet regularization all modes are treated with uniform cutoff.

\section{Conclusions}

In this paper we analyzed in a systematic way the one-loop correction to the energy of a semiclassical 
folded string spinning in \adscp. We derived an integral representation for the energy correction and analyzed its properties for short and long strings. Also, for short strings, we studied the properties of the slope as a first step toward
an eventual exact formula in the spirit of  \cite{Basso:2011rs}. 
In this respect, the main difficulty appears to be 
the inclusion of wrapping effects in Basso's conjecture  since we have shown that these finite size
corrections affect the ABJM slope in contrast to the simpler case of $\mc N=4$ SYM.

\section*{Acknowledgments}

We are especially grateful to Nikolay Gromov for his kind guidance during the course of this project.
We thank Benjamin Basso for important remarks and suggestions.
We also thank Arkady A. Tseytlin for deep and useful  comments. 

\appendix

\section{Redefinition of $g$ and removal of the terms $\sim \log 2$}
\label{app:log2}

Let us show that  terms proportional to $\log 2$ in $E_{1}$ are related to the classical energy by a redefinition of $g$.
To  this aim,  we pick the part of $E_{1}$ proportional to $\log 2$, replace 
\be
\mathcal S = \frac{S}{4\,\pi\,g}, \qquad
\mathcal J = \frac{J}{4\,\pi\,g},
\ee
and expand the coefficients of the various powers of $S$ at large $g$. We find 
\ba
E_{1} &=& \log\,(2)\,\bigg\{
\bigg(\frac{1}{J}-\frac{J}{32 \pi ^2 g^2}+\frac{3 J^3}{2048 \pi ^4 g^4}+\cdots
\bigg)\,S+\\
&& + \bigg(
-\frac{4 \pi  g}{J^3}+\frac{J}{128 \pi ^3 g^3}-\frac{J^3}{1024 \pi ^5 g^5}+\cdots
\bigg)\,S^{2}+\nonumber \\
&&+\bigg(
\frac{24 \pi ^2 g^2}{J^5}+\frac{1}{128 \pi ^2 g^2 J}-\frac{33 J}{2^{13} \pi ^4 g^4}+\frac{775 J^3}{2^{20} \pi ^6
   g^6}+\cdots
\bigg)\,S^{3}+\cdots
\bigg\}+\cdots\nonumber
\ea
On the other hand, we can take $4\,\pi\,g\,\mc E_{0}$ from (\ref{eq:classical}), do the same substitution, set
\be
g\to g+\frac{c}{4\,\pi},
\ee
and expand at large $g$. The part linear in $c$ takes precisely the same form 
\ba
E_{0} = 4\,\pi\,g\,\mc E_{0} &=& c\,\bigg\{
\bigg(\frac{1}{J}-\frac{J}{32 \pi ^2 g^2}+\frac{3 J^3}{2048 \pi ^4 g^4}+\cdots
\bigg)\,S+\\
&& + \bigg(
-\frac{4 \pi  g}{J^3}+\frac{J}{128 \pi ^3 g^3}-\frac{J^3}{1024 \pi ^5 g^5}+\cdots
\bigg)\,S^{2}+\nonumber \\
&&+\bigg(
\frac{24 \pi ^2 g^2}{J^5}+\frac{1}{128 \pi ^2 g^2 J}-\frac{33 J}{2^{13} \pi ^4 g^4}+\frac{775 J^3}{2^{20} \pi ^6
   g^6}+\cdots
\bigg)\,S^{3}+\cdots
\bigg\}+\cdots\nonumber
\ea
This means that choosing 
\be
c = -\log(2)
\ee
removes all $\log 2$ terms from $E_{1}$. 
The above redefinition of $g$ connects the coupling in the algebraic curve regularization with the coupling in the world-sheet regularization and agrees with the calculation in \cite{McLoughlin:2008he} as well as \cite{Abbott:2010yb}.

\bibliography{AC-Biblio}{}
\bibliographystyle{JHEP}

\end{document}